\documentstyle{article}
\begin{document}
\title{Equivalence between Schwinger and Dirac schemes of quantization}
\author{Naohisa Ogawa,\ \ Kanji Fujii,\ \ Hitoshi Miyazaki, \\
Nicolai Chepilko*, \ \ and  Takashi Okazaki**  \\
\\
\ Department of Physics, Hokkaido University, Sapporo 060\\
\   * Institute of Physics, Nauki Prospect 46, Kiev 252650\\
\   ** Physics Department, Hokkaido University of Education, 
Sapporo 002}
\date{January  24,  1996}
\maketitle

\begin{abstract}
 This paper introduces the modified version of Schwinger's quantization 
method, in which the information on constraints and the choice of gauge 
conditions are included implicitly in the choice of variations used in
 quantization scheme. A proof of equivalence between Schwinger- and 
Dirac-methods for constraint systems is given. 
\end{abstract}

\section{Introduction}

\subsection{Purpose}
~~~~ As is well known that the quantization is usually performed on 
the basis of the operator formalism or path-integral formalism. 
In the operator formalism, we replace the Poisson bracket 
(or Dirac bracket \cite{const}) with the commutator. 
Schwinger has proposed another quantization scheme in 
his old papers \cite{schw}, so called the "Schwinger's quantization"
 which is based on the Schwinger's action principle.

In spite of not so clear logical structure, this method has been applied to
 dynamical systems with constraints, such as QED \cite{schw}. The application to 
other gauge theories are given in ref.\cite{ach}.
 Due to such a situation, Das and Scherer have examined the equivalence  
 between the Schwinger and the Dirac schemes of quantization for constraint systems by 
 reinterpreting (or modifying) the Schwinger's quantization scheme\cite{Das}. 
The key point is that they replaced the variation $\delta$ in the Schwinger's 
quantization with another one $\delta_c$, through which the vanishing of 
their brackets among constraints $\phi$'s and anything is automatically insured.
  Note that such a vanishing is a necessary condition for the Dirac bracket. 
 In this sense the bracket has the same property as the Dirac's one, and
 they concluded the Schwinger's quantization scheme is equivalent to the Dirac's one.

 In the present paper, we propose another way of the Schwinger's quantization for
 constraint dynamical systems which is different from the Das-Scherer 
 approach\cite{Das}, and show our quantization scheme to be 
 equivalent to the Dirac one.
In our scheme, modification of the variation $\delta$ is unnecessary;  
instead, we extend the variation $\delta$ to include the q-number variation
 in the Schwinger's scheme, and we restrict $\delta$ so as to satisfy $\delta 
\phi = 0$ automatically for each constraint $\phi$ in a wide class of 
q-number variations. (In refs. \cite{schw} and \cite{Das}, Schwinger as 
well as Das and Scherer employed only c-number variations, so called
 "the fundamental variation"; see ref.\cite{ach} on this point.)
 Our scheme to be described in the following is based on ref.\cite{ach}
  basically, but the treatment for constraints is different and more systematic.

   In the remainder of this section, we introduce the 
Schwinger's quantization scheme with q-number variation,  in which 
variations are functions of dynamical variables.  In section 2, we give a proof of 
equivalence of the Schwinger approach to the Dirac one, where the constraint 
is limited only in the second-class. In section 3 we discuss the case of 
the first-class constraint. In section 4, some examples of 
application of our scheme are given, and in section 5 
a remark is given on the difference of our scheme from that of ref.\cite{Das}.

\subsection{Schwinger's bracket}

~~~~~ For symplicity, we write down all the calculations in quantum mechanics with finite degrees of freedom. 
 As to the field theory, it is enough to remember that each 
contraction of indices is understood as an integration over continuous range. 
 
  Let us consider a dynamical system with a set of  the canonical coordinates $q^k$ 
and  momenta $p_k$, $k=1,2, \cdots $, where the action is given 
as first-ordered form
\begin{equation}
  S = \int dt L(q,p, \dot{q}) = \int dt [\: p_k \dot{q}^k - H(q,p) \:].
\end{equation}
This system may or may not have constraints.
We consider the Lie-variation $\delta$ on the phase space. 
Then we define the charge
\begin{equation}
   G \equiv p_k \delta q^k - (d/dt)^{-1}\delta(p_k \dot{q}^k).
\end{equation}
We can prove, as described in the following paragraph, that $G$ is the Noether charge under the following two conditions;
\begin{description}
  \item[1)]  the variation $\delta$ causes the transformation leaving our action invariant,
  \item[2)]  $\delta q$ and $\delta p$ do not contain time derivatives of canonical variables,\\ and do not include time explicitly.
\end{description}
We note here that the second condition is trivally satisfied when $\delta$
 is a usual canonical transformation;  but this condition holds in a wider 
class of variations.
\\

 The proof of the above statement concerning $G$ is simple. 
From the conditions 1) and 2), we obtain the relation, 
\begin{equation}
  \delta L = \frac{d W}{dt},    
\end{equation}
where $W$ is some function of $q$ and $p$; $W = W(q,p)$.
Due to the condition 2), such a $\delta L$ comes only from the 
kinetic term of the Lagrangian, $L_{kin} = p_i \dot{q}^i$; therefore,
\begin{equation}
  \delta L_{kin} = \delta(p_i \dot{q}^i) = \frac{dW}{dt},\ \ \ \delta H = 0.
\end{equation}
On the other hand,
\begin{equation}
\delta L(q,p,\dot{q}) = \frac{d}{dt}(p_i \delta q^i)-(\dot{p}_i + 
\frac{\partial H}{\partial q^i}) \delta q^i + (\dot{q}^i - 
\frac{\partial H}{\partial p_i})\delta p_i,
\end{equation}
leading to 
\begin{equation}
  \delta L \mid_{on-shell} = \frac{d}{dt}(p_i \delta q^i);
\end{equation}
then the Noether charge Q is given by
\begin{equation}
        Q = p_i \delta q^i -  W = p_i \delta q^i - (\frac{d}{dt})^{-1} 
\delta (p_i \dot{q}^i),
\end{equation}
which means $G$ is the Noether charge for the variation $\delta$.
\

In the following (and also in the following sections) , we confine our considerations to such a variation that does not depend on the time derivatives of $q^i$ and $p_i$;  
\begin{equation}
\delta q^i = f^i(q,p), \ \ \  \delta p_i = g_i(q,p),
\end{equation}
and that may or may not be the symmetry of our action.
The Schwinger bracket is then determined by the following way;
\begin{equation}
\delta q^i \equiv \{ q^i, \ G \}_S,  \ \ \  \delta p_i \equiv 
\{ p_i, \ G \}_S,
\end{equation}
or more precisely, the fundamental brackets between canonical 
variables $(q,p)$ are determined in accordance with
\begin{eqnarray}
\delta q^i & \equiv & \{ q^i, \ q^j \}_S \frac{\partial G}{\partial q^j} 
+  \{ q^i, \ p_j \}_S \frac{\partial G}{\partial p_j}, \\
\delta p_i & \equiv & \{ p_i, \ q^j \}_S \frac{\partial G}{\partial q^j} 
+  \{ p_i, \ p_j \}_S \frac{\partial G}{\partial p_j}.
\end{eqnarray}
To obtain all the fundamental brackets, it might be necessary to utilize
 a certain number of  variations, and the Jacobi identity can also be helpful.

     Let us consider the condition for all the fundamental brackets to be 
determined.  For this  purpose, we denote coordinates simply as
\begin{equation}
              \{z^i\} = \{q, p \}, \ \  (i=1 \sim 2N)
\end{equation}
where 2N is the dimension of the phase space.  We consider the case,
 in which there are n $(n \ge 2N)$ independent generators $G^{\alpha}  
\ \ (\alpha = 1 \sim n) $  , and each $G^{\alpha}$ contains only one 
variational parameter $\varepsilon_{\alpha}$.  Then we have 
 \begin{equation}
   \delta^{\alpha} z^k = \varepsilon_{\alpha}  f^{\alpha}_k(z).
 \end{equation}
 Note that we do not take a contraction for index $\alpha $ when no  
 summation symbol is accompanied.  From (10) and (11) we have 
 \begin{equation}
 \delta^{\alpha} z^k = \sum_j \{z^k, z^j\}_S  \  \frac{\partial 
   G^{\alpha}}{\partial z^j}.
 \end{equation}
 Since $G$ is a linear and homogeneous function of $\delta$ due to (2), 
 we can write as
 \begin{equation}
     \frac{\partial G^{\alpha}}{\partial z^j} = M^{~\alpha}_j(z)  
     \ \varepsilon_{\alpha}.
 \end{equation}
 Then we obtain the equation as,
 \begin{equation}
   \sum_{j} \{z^k, z^j\}_{S} \ M^{~\alpha}_j = f^{\alpha}_{k}.
 \end{equation}
   The condition for all the fundamental brackets to be 
   determined is the existence of the (right) inverse matrix of $M$;  
then we obtain

   \begin{equation}
   \{z^k, z^j\}_S = \sum_{\alpha} f^{\alpha}_{k} \ (M^{-1})_{\alpha}^{~j}.
   \end{equation}
         
The quantization is performed by replacing the Schwinger bracket with
 the commutation relation as
\begin{equation}
   \{ A, \  B \}_S \  \longrightarrow \  \frac{1}{i \hbar} [ A, \ B ].
\end{equation}
 Applications of this Schwinger's quantization are given in ref. 
\cite{ach}.
\

\section{Theorems for equivalence of brackets}
~~~ ~~~ Now we can prove the following theorems.
\

\

Theorem 1: \  In a non-singular system, if we find the local form of 
$K(q,p)$ which satisfies  $\delta(p_i \dot{q}^i) = dK/dt$, 
it is proved that
\begin{equation}
   \{ A, \ B \}_S \ = \ \{ A, \ B \}_P 
\end{equation}
holds under the condition for all the fundamental Schwinger brackets to be determined, where $P$ denotes Poisson bracket. 
\

\

 The proof is given as follows.
From the condition of the theorem,
\begin{equation}
\delta(p_i \dot{q}^i) = \frac{dK}{dt}
\end{equation}
holds, and from this equation we obtain
\begin{equation}
(\delta p_i + p_j \frac{\partial (\delta q^j)}{\partial q^i}) 
\dot{q}^i + p_j \frac{\partial (\delta q^j)}{\partial p_i} \dot{p}_i 
= \frac{\partial K}{\partial q^i} \dot{q}^i + \frac{\partial K}
{\partial p_i} \dot{p}_i.
\end{equation}
Therefore, from the independence of the variables  $(\dot{q}, \dot{p})$, 
 K should satisfy the conditions
\begin{eqnarray}
\frac{\partial K}{\partial q^i} \: &=& \: \delta p_i + p_j 
\frac{\partial (\delta q^j)}{\partial q^i},\\ 
\frac{\partial K}{\partial p_i} \: &=& \: p_j \frac{\partial 
(\delta q^j)}{\partial p_i}. 
\end{eqnarray}
Thus, from the definition of G,  i.e.  $ G \equiv p_i \delta q^i - K$, we obtain
\begin{eqnarray}
\frac{\partial G}{\partial q^i} \: &=& \: \frac{\partial}{\partial q^i}
 (p_j \delta q^j - K) = - \delta p_i,\\ 
\frac{\partial G}{\partial p_i} \: &=& \: \frac{\partial}{\partial p_i}
 (p_j \delta q^j - K) = \delta q^i. 
\end{eqnarray}
Then from the definition of the Schwinger bracket, we obtain the first result
\begin{eqnarray}
\{ q^i, \ G \}_S &=& \: \: \frac{\partial G}{\partial p_i} = \{ q^i, 
\ G \}_P,\\
\{ p_i, \ G \}_S &=& - \frac{\partial G}{\partial q^i} = \{ p^i, \ G \}_P.
\end{eqnarray}
 
Since the original definition of the Schwinger bracket is given as (9), 
 we can not obtain further relations formally.  But when all the 
 fundamental Schwinger brackets are determined as we have discussed 
 at the end of section 1, we can prove that all the fundamental 
 brackets are the same as Poisson brackets.
 Because,  firstly  we can rewrite the above relations by using (15) in the form
 \begin{equation}
   0 = \sum_{j} [ \{z^i, z^j\}_S  -  \{z^i, z^j\}_P ]\frac{\partial G^{\alpha}}
   {\partial z^j} = \sum_j [ \{z^i, z^j\}_S  -  \{z^i, z^j\}_P ] 
   \ M^{~\alpha}_j \varepsilon_{\alpha}.
 \end{equation}
 
 Then from the  condition for determing the fundamental bracket, 
 that is, the existence of inverse of $M$,  we obtain 
 \begin{equation}
 \{z^i, z^j\}_S  =  \{z^i, z^j\}_P.
 \end{equation} 
 Therefore, the Schwinger bracket coincides with the Poisson bracket.
\

\

Nextly we work with the second-class constraint system; then the 
theorem can be extended as follows.
\

\

Theorem 2: \ In a singular system, if we find the local form of 
$K(q,p)$ which satisfies  $\delta(p_i \dot{q}^i) = dK/dt$, 
and if $\delta$ satisfies $\delta \phi_m = 0 $
 for all constraints $\phi_m$'s,  it is proved that  
\begin{equation}
   \{ A, \ B \}_S \ = \ \{ A, \ B \}_D 
\end{equation}
holds under the condition for all the fundamental Schwinger brackets 
to be determined, where $D$ denotes Dirac bracket. 
\

\

The proof is coming here.
In the singular system, because of existence of the constraints:
\begin{equation}
          \phi_m (q,p) \ = \ 0,
\end{equation}
q's and p's are no longer independent variables. So we must rewrite equation (21)
 by adding the relations
\begin{equation}
 \frac{\partial \phi_m}{\partial q^i} \dot{q}^i + \frac{\partial \phi_m}
{\partial p_i} \dot{p}_i = 0
\end{equation}
with arbitrary coefficients $\lambda^m$.  Thus the extended versions of
 (22) and (23) are expressed as
\begin{eqnarray}
\frac{\partial K}{\partial q^i} \: &=& \: \delta p_i + p_j \frac{\partial
 (\delta q^j)}{\partial q^i} + \lambda^m \frac{\partial \phi_m}
{\partial q^i},\\ 
\frac{\partial K}{\partial p_i} \: &=& \: p_j \frac{\partial 
(\delta q^j)}{\partial p_i} + \lambda^m \frac{\partial \phi_m}
{\partial p_i}. 
\end{eqnarray}

 From the definition of G, we obtain
\begin{eqnarray}
\frac{\partial G}{\partial q^i} \: &=& \: \frac{\partial}{\partial q^i}
 (p_j \delta q^j - K) = - \delta p_i - \lambda^m \frac{\partial \phi_m}
{\partial q^i},\\
\frac{\partial G}{\partial p_i} \: &=& \: \frac{\partial}{\partial p_i}
 (p_j \delta q^j - K) = \delta q^i - \lambda^m \frac{\partial \phi_m}
{\partial p_i},
\end{eqnarray}

from which we have
\begin{eqnarray}
\{ q^i, \ G \}_S &=&  \{ q^i, \ G \}_P + \lambda^m \frac{\partial \phi_m}
{\partial p_i},\\ 
\{ p_i, \ G \}_S &=&  \{ p^i, \ G \}_P - \lambda^m \frac{\partial \phi_m}
{\partial q^i}.
\end{eqnarray}
From the requirement 
\begin{equation}
\delta \phi_m(q,p) = 0  \ \ \ (for~all~\phi_m~'s), \ \ \ 
\end{equation}
we have
\begin{eqnarray}
\delta \phi_m &=&  \frac{\partial \phi_m}{\partial p_i} \delta p_i + 
\frac{\partial \phi_m}{\partial q^i} \delta q^i,\\
&=&  \frac{\partial \phi_m}{\partial p_i} \{ p_i, G \}_S +
 \frac{\partial \phi_m}{\partial q^i} \{ q^i, G \}_S = 0.
\end{eqnarray}
Putting (37) and (38) into (41) , we obtain
\begin{equation}
\{ \phi_m, \ G \}_P  +  \lambda^k \{ \phi_m, \ \phi_k \}_P = 0.
\end{equation}
These equations completely determine $\lambda$'s when the constraints
 form the 
second-class, and
\begin{equation}
   \lambda^m = - (u^{-1})^{mn} \{ \phi_n, \ G \}_P; \ \ \  u_{mn} 
\equiv \{ \phi_m, \ \phi_n \}_P.
\end{equation}
From  (37), (38) and (43), we obtain for any function $X(q,p)$ that
\begin{eqnarray}
\{ X(q,p), G \}_S &=& \{ X(q,p), G \}_P - \{ X(q,p), \phi_m \}_P
 (u^{-1})^{mn} \{ \phi_n, G \}_P \nonumber\\
  & = & \{ X(q,p), G \}_D.
\end{eqnarray}
In the same way as the explanation at the end of the proof of Theorem1, we 
can prove the equivalence between the fundamental Schwinger  and 
Dirac brackets by using the above equality. 
In the gauge constraint system we meet, however, the case, in which  we can 
not prepare enough variation parameters, $n$ in $G$. In this case, some 
of the Schwinger brackets can be determined to be equal to Dirac brackets 
due to (44), by expanding $G$ in both hand sides  with respect to the variation parameters $\varepsilon_\alpha$  and comparing the coefficients 
of each $\varepsilon_\alpha$, but others remain still undetermined. 
We discuss such cases in section 4.

Note that the requirement $\delta \phi = 0$ is essential for our 
constraint system, and $\delta$ includes the information of constraints
 implicitly.
Further if we consider the variations which break the constraint
 conditions,  the constraints are not consistent with quantization
 and have no sense in that dynamical system. It is possible to extend 
the theorems described above so as to include fermions generally.
\

\section{1-st class constraint system}
~~~~~~ In the case of the second-class constraint system, 
 by restricting the variation $\delta$  to (39),  our Schwinger brackets
 coincide with Dirac's ones. In other words,  if we choose $\delta$ 
to calculate  the Schwinger bracket, 
the constraint conditions are implicitly selected. From this fact,  the 
Schwinger bracket can be constructed also in the first-class system.
 Let us denote $\{ \phi_m \}, (m = 1 \sim N)$ as a set of the first-class 
constraints, and $ \{ \psi_n \}, (n = N+1 \sim 2N)$ as gauge conditions.
 We write the total set as  $ \Phi_{\mu} = \{ \phi_m, \psi_n \}, 
 (\mu = 1 \sim 2N)$.  For constructing the Schwinger bracket, 
we do not touch the Hamiltonian of the system, 
but we choose the variation $\delta$ so as to satisfy
\begin{equation}
            \delta \Phi_{\mu} = 0,  ~~~~~ \mu =  1 \sim 2N.
\end{equation}
Putting it in another way, suppose we choose some favorite $\delta$,
 where the above conditions are accidentally satisfied; and if 
\begin{equation}
           det \{ \Phi_{\mu},  \Phi_{\nu} \}_P \neq 0
\end{equation}
is satisfied, then the Schwinger bracket can be obtained in the same way 
as the second-class case (Theorem 2). If the condition (46) does not hold,
 some $\lambda$'s  remain undetermined in (43), and the Schwinger bracket
 can not be  fully calculated.
The obtained Schwinger brackets in this way are exactly the same as the 
Dirac bracket with constraints $\Phi_{\mu} = 0$. On the choice of $\delta$, it
 would be helpful to comment the following point.  There are two conditions which
 $\delta$ should satisfy. The first condition, existence of $K(q,p)$, is 
satisfied when $\delta$ is the symmetry of our system. The second condition
 is $\delta \Phi = 0$, which means that the gauge constraints do not break
 the symmetry. Therefore $\delta$ is the resultant symmetry after the gauge
 fixing. The examples are, BRS symmetry, global gauge symmetry, Poincare
 symmetry  and other trivial ones.

\section{Examples}
~~~~~~ The quantization procedure is, however, not yet 
finished at this stage.
 Let us denote that all the canonical variables $ Z= \{q, p\} =
 \{z, \bar{z} \}$, where some of variables $\bar{z}$ are included
 in the charge $G= G(\bar{z})$. 
From the quantization scheme expressed in the previous section, we can
 determin almost the Schwinger brackets with form:
\begin{equation}
           \{ z_i, \bar{z}_j \}_S,  \{\bar{z}_i, \bar{z}_j \}_S,
\end{equation}
but not others. On the rewriteness from brackets to commutation relations,
 we naturally require the symmetry (antisymmetry for fermion case)  when interchanging the variables in the bracket. So we can also obtain
\begin{equation}
            \{ \bar{z}_i, z_j \}_S.
\end{equation}
Now the completely undetermined parts have the form $\{z, z\}_S$.  The most
 suitable way to determine this part and other undetermined brackets is
 to require the necessary condition for the Dirac bracket:
\begin{equation}
           \{ \phi_m, Z \}_S = 0, ~~~~ m=1,2, \cdots , 
\end{equation}
We can not say that these are the sufficient conditions for all the 
fundamental brackets to be determined; however, in almost all the
 theories the condition (49) is sufficient.  And from the theorem, if all
 the fundamental brackets are determined in this way, they are completely
 equivalent to the Dirac brackets.
We will show such a situation in two examples; a relativistic free particle
 and the pure QED.

\

\
(1) Relativistic free particle 
\

Let us work with the following action,
\begin{equation}
 L = p_{\mu} \dot{x}^{\mu} - N(p^2 - m^2).
\end{equation}
There are two first-class constraints;
\begin{equation}
   \phi_1 = \pi_N, \:\:\:  \phi_2 = p^2 - m^2.
\end{equation}
We also add here the gauge conditions;
\begin{equation}
   \phi_3 = N, \:\:\:  \phi_4 = x^0 - t.
\end{equation}
Then the variation in Schwinger's quantization scheme should satisfy
\begin{eqnarray}
(1)  && \delta (p_{\mu} \dot{x}^{\mu} + \pi_N \dot{N}) = \dot{K} \; 
\;(for ~some ~local ~form~ K),\\
(2)  && \delta \phi_{\mu} = 0. \;\; (\mu = 1 \sim 4).
\end{eqnarray}
The general solution for these two conditions is the following;
\begin{eqnarray}
 && \delta p_0 = 0, \:\: \delta x^0 = 0, \:\: \delta \pi_N = 0, \:\: 
\delta N = 0,\\
 && \delta p_i = \epsilon_{ijk} \beta^j p^k, \:\: \delta x^i = 
\epsilon^i_{\: jk} \beta^j x^k + \lambda p^i + \mu^i,
\end{eqnarray}
where $\beta, \: \lambda, \: \mu$ are the constant parameters,  
and $i,j,k = 1 \sim 3$.  The related  "charge" G has the form
\begin{equation}
       G = -\frac{1}{2} \lambda \vec p^2 + \vec \beta \cdot (\vec x \times
 \vec p) + p_i \mu^i.
\end{equation}
We can easily find that the parameters $\lambda$ , $\beta$ and $\mu$ 
correspond to the energy conservation,  to the angular momentum 
conservation, and to the momentum conservation, respectively.
 The other symmetries are broken by the constraints. By using above G and
 $\delta$, we obtain the following Schwinger brackets;
\begin{eqnarray}
&& \{x^i, p_j\}_S = \delta^i_j, \; \; \{x^i, x^j\}_S = \{p_i, p_j\}_S = 0,
 \nonumber\\
&& \{p_0, p_i\}_S = \{x^0, p_i\}_S = \{N, p_i\}_S = \{\pi_N, p_i\}_S = 0,
\nonumber\\
&& \{p_0, x^i\}_S = A p^i, \:\: \{x^0, x^i\}_S = B p^i,\nonumber\\
&& \:\: \{N, x^i\}_S = C p^i, \:\: \{\pi_N, x^i\}_S = D p^i,\nonumber
\end{eqnarray}
where $A,B,C,D$ are undetermined quantities. Let us require the condition
 $\{\phi , x^i \}_S = 0$ for the brackets. Then we immediately obtain
 $B=C=D=0$, and from $\{ p^2, x^i \}_S = 0$, $A=1 / p_0$ is obtained.
 These obtained brackets are equivalent to the Dirac ones, and other
 $\{z,z\}_S$ type brackets are also determined to be 0 in the same way. 
We see that G is really the resultant 
symmetry of our theory  after gauge fixing.

\

\

(2) Pure QED
\

We start from the first-ordered Lagrangian density for QED in the form
\begin{equation}
 L = \pi^k \dot{A}_k + \frac{1}{2} \pi^k \pi_k - \frac{1}{4} F_{ij} F^{ij}
 + A_0 \partial_k \pi^k,
\end{equation}
where we utilize the metric $\eta = (1,-1,-1,-1)$, and we identify $A_0$
 as the multiplier field but not as the coordinate. Then we obtain only one
 first-class constraint
\begin{equation}
\phi = \partial_k \pi^k .
\end{equation}
As the related gauge condition, we take the Coulomb gauge
\begin{equation}
\phi_g = \partial_k A^k .
\end{equation}

~~~~~~Let us choose the variation in the following form;
\begin{equation}
\delta A^i(x) = g^i(x), \ \ \ \delta \pi^i(x) = h^i(x),
\end{equation}
where $g$ and $h$ are some functions which do not depend on dynamical
 variables. By requiring the condition that $K$ has a local form, we find
 that $g$ and $h$ are time-independent functions, and then $K$ and $G$ 
 are expressed as 
\begin{equation}
     K = \int dx \: h_i A^i,\ \ \ G = \int dx [\pi_k g^k - h_k A^k].  
\end{equation}

Nextly the condition that the constraint is invariant under the variation 
requires 
\begin{equation}
\delta \phi = \partial_k h^k = 0, \ \ \  \delta \phi_g = \partial_k g^k =0.
\end{equation}

The Schwinger brackets are determined by $g^i = \{A^i, G\}_S$ and $h^i =
 \{\pi^i, G\}_S$, which lead to
\begin{eqnarray}
g^i(x) &=& \int dy \{A^i(x), \pi_k(y)\}_S \: g^k(y) -\int dy \{A^i(x),
 A_k(y)\}_S \: h^k(y),\\
h^i(x) &=& \int dy \{\pi^i(x), \pi_k(y)\}_S \: g^k(y) -\int dy 
\{\pi^i(x),A_k(y)\}_S \: h^k(y).
\end{eqnarray}
Then we obtain the fundamental brackets with arbitrary functions $N,F,S$ as
\begin{eqnarray}
\{A_i(x), \pi_j(y)\}_S &=& \eta_{ij} \delta(x-y) + \partial_i^x 
\partial_j^y N(x,y),\\
\{A_i(x), A_j(y)\}_S &=& \partial_i^x \partial_j^y F(x,y),\\
\{\pi_i(x), \pi_j(y)\}_S &=& \partial_i^x \partial_j^y S(x,y),
\end{eqnarray}
where  arbitrary functions $N, F $ and $S$ are coming from the condition $\partial_i g^i = \partial_i h^i = 0$. We can determine these arbitrary functions
 by requiring that the brackets should vanish if the constraints appear
 inside, that is, if we take the derivative $\partial_i^x$ or 
$\partial_j^y$ for the above equations, the left hand sides vanish.  Also
 from the fact that the solution of $\nabla^2 F(x) = 0$ damping  at 
infinity is only $F(x) = 0$, we find that $F(x,y) = S(x,y) =0$. For 
$N(x,y)$ we obtain
\begin{equation}
       \nabla^2_x N(x,y) = \nabla^2_y N(x,y) = \delta(x-y).
\end{equation}
The solution is given as
\begin{equation}
  N(x,y) = - \frac{1}{(2\pi)^3} \int d^3p \: \frac{
e^{i\vec{p} \cdot (\vec{x}-\vec{y})}}{\vec{p}^2}.
\end{equation}
Therefore, the final result is given as
\begin{eqnarray}
\{A_i(x), A_j(y)\}_S &=& \{\pi_i(x), \pi_j(y)\}_S = 0,\\
\{A_i(x), \pi_j(y)\}_S &=& \frac{1}{(2\pi)^3} \int d^3p \: \frac{p_ip_j 
- {\vec{p}}^2 \delta_{ij}}{\vec{p}^2} e^{i\vec{p} \cdot (\vec{x}-\vec{y})},
\end{eqnarray}
which are the usual commutation relations in the Coulomb gauge.

\section{Discussion}
~~~~  We give some remarks on the difference between the  method given
 by  Das and Scherer \cite{Das} and ours.
Firstly the definition of  the "Schwinger's quantization scheme" is somewhat
 different with each other. If we restrict the variations in our scheme to the
 following two ways;
\begin{eqnarray}
&1:&  \delta q = \xi = const, \;\; \delta p = 0.\\
&2:&  \delta q = 0, \;\; \delta p = \eta = const.
\end{eqnarray}
then these are the ones utilized in \cite{Das}. In their paper they have
 extended the definition of the Schwinger bracket for the constraint systems:
\begin{eqnarray}
&& \delta f \;\; \rightarrow \;\; \delta_c f = \delta f + \Lambda_f^i
 \delta \phi_i,\\
&& and, \;\; \delta_c f \equiv \{ f, G \}_S.
\end{eqnarray}
The condition $ \{\phi, f(q,p) \}_S = 0 $ is satisfied by choosing the
 multiplier field $\Lambda$. Then its form is determined, which is seen
 to be equal to the Dirac bracket.  While, in our case our choice of 
$\delta$ has more varieties,
 and we restrict them so as to satisfy $\delta \phi = 0$ instead of changing
 the definition of variation.  In the Das-Scherer method, they must 
calculate the multiplier field  to obtain the bracket which is the same 
task to calculate the Dirac bracket.  On the other hand in our method, we
 should look for the resultant symmetry $\delta$ related to constraints.
\newpage

\noindent{\em Acknowledgement.}\\
The authors would like to thank Prof. R. Jackiw for pointing out 
a exceptation for our proof, on which we have improved in this new version.

\end{document}